# Low-Cost Shield MicroBCI to Measure EEG with STM32

Ildar Rakhmatulin, PhD

**Abstract**
The article introduces an accessible pathway into neuroscience using the MicroBCI device, which leverages the STM32 Nucleo-55RG development board as the core platform. MicroBCI enables the STM32 board to function as a brain–computer interface, capable of recording EEG, EMG, and ECG signals across 8 channels. Over the past decade, the rapid growth of artificial intelligence has transformed many fields, including neurobiology. The application of machine learning methods has created opportunities for the practical use of EEG signals in diverse technological domains. This growing interest has fueled the popularity of affordable brain–computer interface systems that utilize non-invasive electrodes for EEG acquisition. The MicroBCI device demonstrates reliable noise performance and accuracy for applied research and prototyping. Furthermore, it effectively detects alpha brain waves, confirming its ability to capture key neurological signals.

Source in GitHub  https://github.com/pieeg-club/MicroBCI

## 1. Introduction

Electroencephalography (EEG) is a widely recognized and accessible method for recording brain activity. It functions by measuring the electrical signals generated in different brain regions using electrodes placed on the scalp. Due to its versatility, EEG has become a fundamental tool in both research and clinical practice, with applications that continue to expand. A variety of electrode types are available for signal acquisition, most notably wet and dry electrodes. STM32 is one of the most popular microcontrollers on the market. However, to the best of our knowledge, there are no boards available that provide a convenient format along with an SDK and mobile SDK for turning STM32 boards into a brain computer interface.

## 2. Problem Statement and Motivation

The price of commercial brain–computer interface (BCI) systems varies considerably, with many devices remaining too costly for researchers, educators, and hobbyists. Meanwhile, the rapid progress of neural networks and signal processing techniques has sparked a surge of interest in BCIs across multiple application domains. As early as 2016, Meng et al. [2] demonstrated the control of a robotic arm using motor imagery, where neural networks decoded brain signals collected through electrodes. Previously, such outcomes were mostly limited to invasive methods employing implanted microelectrode arrays to record brain activity directly from the cortex [3]. Although effective, invasive approaches are expensive, technically challenging, and carry significant risks, requiring specialized personnel and equipment [4].

This has driven increasing attention toward non-invasive, low-cost EEG acquisition platforms. Conventional EEG readers are typically composed of a microcontroller (processor), a power supply board, and communication modules—elements that often contribute to their high cost. To mitigate this issue, we developed the microBCI shield for the STM32 Nucleo-55RG board, which extends the capabilities of the Nucleo platform to acquire biosignals such as EEG, EMG, and ECG. Leveraging the Nucleo ecosystem's popularity for embedded systems education and prototyping, this approach lowers the entry barrier and offers an accessible solution for affordable biosignal measurement.

## 3. Review of Related Work

A wide range of devices has been developed for recording EEG signals, each with varying levels of accessibility and technical sophistication. Gunawan et al. [5] employed a low-cost system for EEG signal acquisition aimed at classification tasks, while Ashby et al. [6] used a similar setup to classify mental visual activities. A general overview of brain computer interfaces is presented in the paper [7]. Building upon these efforts, this article introduces the microBCI shield, a self-contained, low-cost biosignal interface designed for the STM32 Nucleo-55RG board. This approach provides a practical and extensible platform for EEG, EMG, and ECG acquisition, while leveraging the robustness and ecosystem support of the STM32 Nucleo family.

## 4. Technical Realizations

In EEG signal acquisition, the ADS1299 analog-to-digital converter (ADC) from Texas Instruments plays a central role. Having been on the market for over a decade, it is widely recognized as one of the most reliable and high-performance ADCs for biopotential measurements. A key advantage of the ADS1299 is its integrated multiplexer, which simplifies multi-channel EEG acquisition. The capabilities of this ADC, along with the features of its multiplexer, have been extensively reviewed by Rashid et al. [8], though a detailed discussion is beyond the scope of this article. Wet electrodes are commonly preferred because of their relatively low impedance, typically ranging between 200 kΩ and 5 kΩ after applying conductive gel. Li et al. [1] provided a comprehensive comparison of wet and dry electrodes, outlining the specific strengths and limitations of each. In our study, we employed dry Ag/AgCl electrodes, as they eliminate the need for gel application and thereby simplify the EEG recording process.

For signal acquisition, we placed 8 dry Ag/AgCl electrodes according to the International 10–20 Electrode Placement System at the following positions: F7, Fz, F8, C3, C4, T5, Pz, and T6. A general overview of the microBCI shield for STM32 Nucleo-55RG is presented in Figure 1.

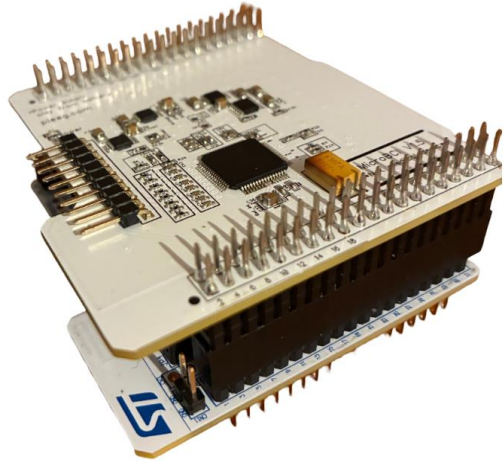

Fig. 1. General view of the PCB boards with soldered elements with STM32 Nucleo-55RG

During testing, the MicroBCI shield operated completely offline, disconnected from any Power source. A power bank was used to ensure safety and prevent potential interference from external connections. The device was powered solely by a 5V battery supply, and it should always be tested and operated using 5V batteries only. A general overview of the assembled device configuration and the electrode placement is shown in Figure 2.

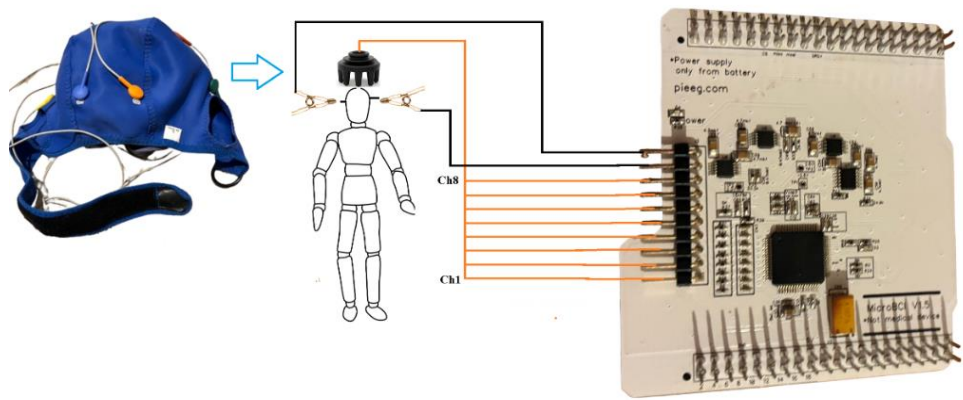

Fig. 2. General view for connecting EEG electrodes to MicroBCI

## 5. Dataset Evaluation

### 5.1. Artefacts

The internal noise of the microBCI shield is approximately 1 µV (dataset available at https://github.com/Ildaron/ardEEG/tree/main/dataset/internal_noise). We conducted several tests to record EEG signals and identify common artifacts, such as chewing and blinking, which are widely used for evaluating EEG device performance. In our experiments, chewing artifacts were clearly distinguishable from the background EEG signal (Figure 3).

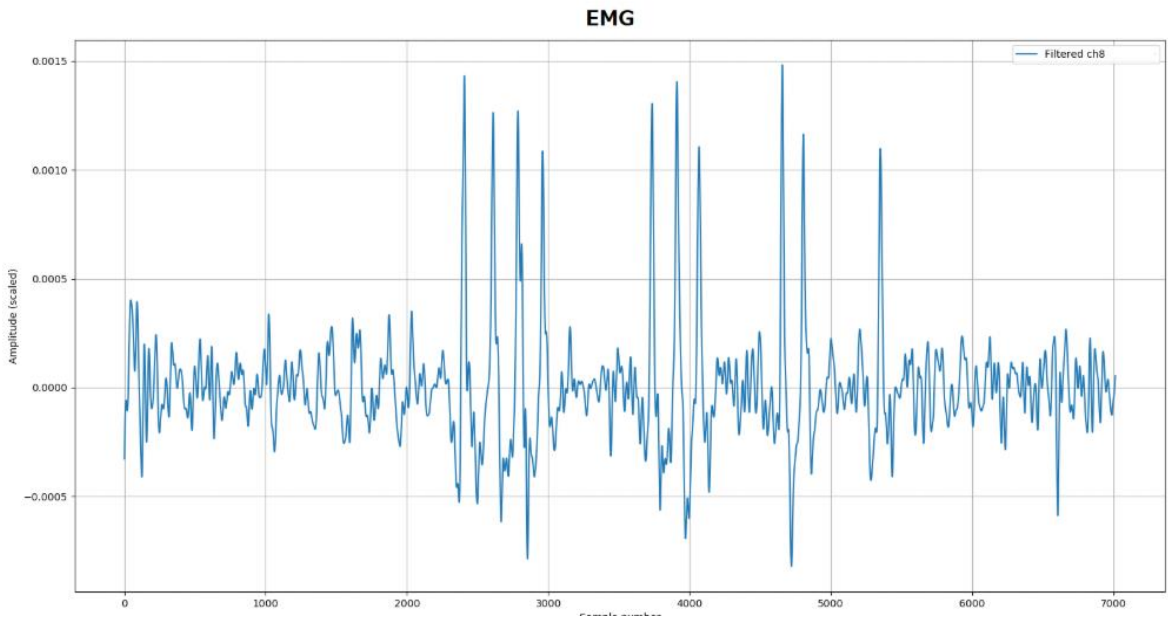

Fig. 3. Artifact test. The process of measuring chewing artifacts using dry electrodes (Fz). Chewing occurred in the following sequence: 4 times, 3 times, 2, and 1 time. The y-axis is the processed EEG signal after passing filter bands of 1-40 Hz in microvolts and with 250 samples per second

## 5.2. Alpha

Alpha waves (α-waves), with frequencies ranging from 8 to 12 Hz and typical amplitudes between 35 and 65 µV, provide a standard benchmark for evaluating EEG recording systems. These waves are usually observed in individuals who are awake, relaxed, and have their eyes closed. In our tests, EEG signals were recorded for 8-second intervals under both eyes-closed and eyes-open conditions. As expected, an increase in EEG amplitude was observed within the 8–12 Hz frequency range when the eyes were closed, while alpha activity decreased when the eyes were open. These results are consistent with the characteristic alpha wave pattern in the occipital region, confirming the proper functionality and design of the microBCI shield (Figure 4).

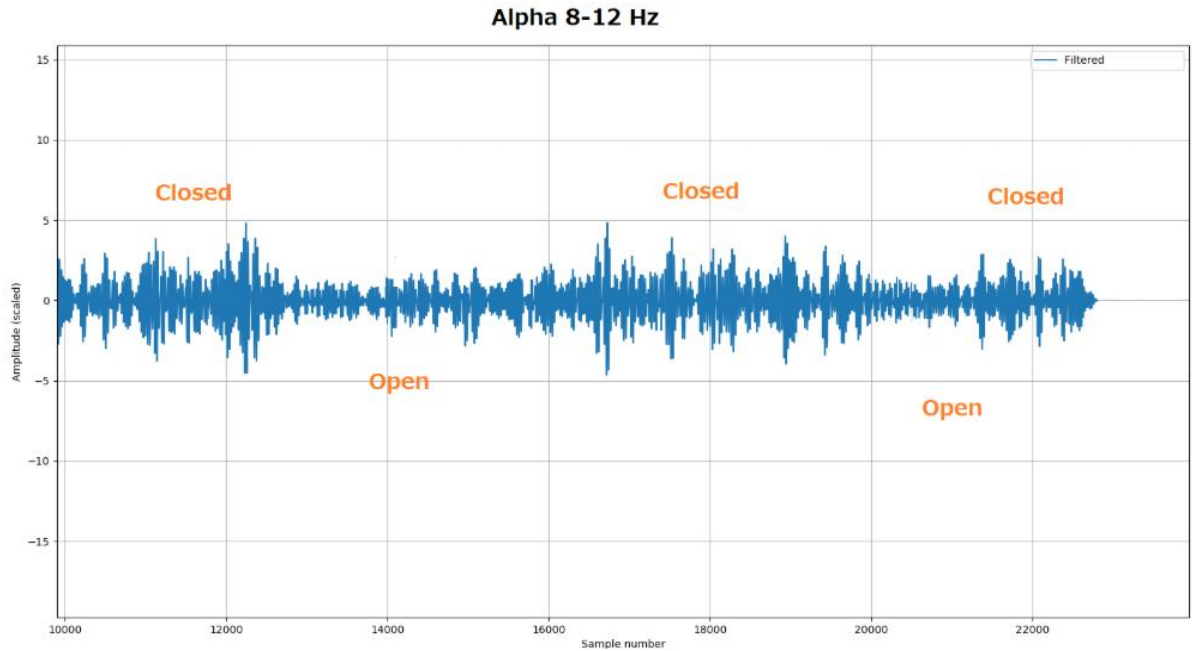

Fig. 4. Alpha test. The process of recording an EEG signal from an electrode (Fz) with eyes open and closed. The y-axis is the processed EEG signal after passing filter bands of 8-12Hz in microvolts, and with 250 samples per second

**6. Conclusion and Discussion**

This article highlights the microBCI shield as a cost-effective yet high-performance solution for EEG, EMG, and ECG signal acquisition. Its accuracy was validated through standard EEG benchmarks, including common artifacts and alpha wave detection, demonstrating the reliability of the device in capturing neurological signals. The shield's noise characteristics closely match those of the ADS1299 ADC from Texas Instruments, confirming high-quality signal acquisition at a fraction of the cost. Efficient data transfer between the ADC and the Nucleo processor ensures real-time operation, making the device suitable for dynamic brain–computer interface (BCI) applications. Integration with the STM32 Nucleo ecosystem provides an accessible platform for researchers, educators, and developers to prototype, experiment, and innovate in BCI technology. By providing the hardware design and software in an open-source format, this project encourages the creation of large, collaborative EEG datasets and facilitates community-driven development in biosignal processing and neurotechnology research. The microBCI shield thus represents a versatile foundation for advancing applied BCI solutions.